\documentclass{ws-p10x7}
\usepackage{graphicx}

\begin{document}

\title{The Muon Anomaly: Experiment and Theory}

\author{J. P. Miller}

\address{for the Muon (g-2) Collaboration\cite{mug2}}

\address{Department of Physics, Boston University, 590 Commonwealth Ave.,
Boston, MA 02215
USA\\E-mail: miller@bu.edu}

\twocolumn[\maketitle\abstract{
A summary is given of the present status of the theory and experiment
of the anomalous magnetic moment of the muon. A difference
between predicted and measured values is an indication
of physics beyond the Standard Model.
A new experimental
measurement has produced a value which differs from a recent
Standard Model prediction by about 1.6 standard deviations.
When first announced, the discrepancy was about 2.6 standard deviations, but
theorists have recently found an error in the sign of the largest term in the
standard model hadronic light-by-light contribution which reduces the
difference.
Additional data are being analyzed and elements of the theory are
being scrutinized to provide, in the future, a sharper test of theory.}]

\section{Introduction}

The magnetic moment of a particle is given in terms of its spin by

$${\vec \mu}={ge \over 2mc}{\vec s}$$

\noindent
and the anomaly is defined as $a={1 \over 2}(g-2)$.

Historically, the measurement of particle magnetic moments has been a
valuable test of existing theories. For example,
magnetic moment measurements
on the hyperons have provided essential
information on their substructure, and the electron anomaly
has been the most stringent test of QED.

The Dirac theory predicts that $g=2$ ($a=0$)
for point particles with spin
${1 \over 2}$. While the hyperons have g factors very different from 2
because of their complex substructure, the leptons have
$g\approx 2$ and anomalies which are nearly zero, consistent with
the current evidence that they are point particles.
The Standard Model predicts lepton anomalies on the order of one
part in 800 due to their field interactions.
In the cases of the electron and the muon,
both the Standard Model predictions and the measurements
are extremely precise, a relatively rare situation
resulting in valuable tests of the theory. 
As we shall see, however, the muon typically has a much stronger sensitivity than
the electron to any physics which has not been included in the Standard Model.

\begin{figure}
\epsfxsize120pt
\includegraphics[angle=0,width=.45\textwidth]
{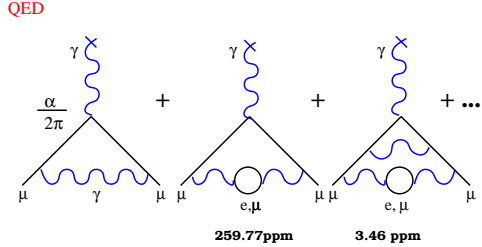}
\caption{QED contributions to the anomaly. The first diagram is the
lowest-order Schwinger term.
The other diagrams are representative of higher-order QED contributions.}
\label{fig:qed}
\end{figure}

\begin{figure}
\epsfxsize120pt
\includegraphics[angle=0,width=.2\textwidth]
{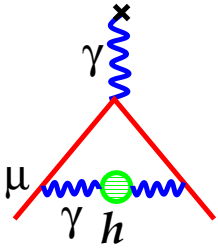}
\caption{The first-order hadronic diagram.}
\label{fig:had1}
\end{figure}

By far the largest contribution to the lepton
anomalies comes from the lowest order
electromagnetic diagram, the Schwinger term
(left diagram in Fig.~\ref{fig:qed}), which gives
$a(QED;1)={\alpha \over 2\pi}$, e.g. the same for muons and electrons.
The next order electromagnetic diagrams,
which involve virtual lepton (Fig.~\ref{fig:qed}) or hadron
(Fig.~\ref{fig:had1}) loops,
are small compared to the Schwinger term, however, they are much larger
for the muon than the electron as a result of
the additional available rest mass
energy. The difference in the contribution between
the electron and muon in diagrams involving
massive virtual particles
typically scales as $({m_{\mu} \over m_e})^2\approx 40000$, and it is this
large factor which makes $a_{\mu}$ far more sensitive than $a_e$
to any unknown massive particles.

It is instructive to compare the values of the electron and muon anomalies.
From Penning trap\cite{dehmelt} experiments, we have for the electron
$a_{e-}^{exp}=(1159652.1884\pm.0043)\times 10^{-9}$ (4 ppb) and positron
$a_{e+}^{exp}=(1159652.1879\pm.0043)\times 10^{-9}$ (4 ppb), agreeing within 
errors as required by CPT invariance.
These values are  are in good agreement with the theoretical prediction
of QED to fourth order in
$\alpha \over \pi$,
$a_e^{th}=(1159652.1535\pm.0240)\times 10^{-9}$ (21 ppb)\cite{CM}.
The theoretical error is dominated by the error in the value
of $\alpha$ taken from Quantum Hall Effect experiments.
If one assumes the correctness of QED, then the best
determination of $\alpha$ comes from the electron anomaly measurements.
The hadronic and electroweak contributions,
$1.63(3)\times 10^{-12}$ and $0.030 \times 10^{-12}$, respectively,
are small and the errors are negligible compared to the experimental errors.

\begin{figure}
\epsfxsize120pt
\includegraphics[angle=0,width=.48\textwidth]
{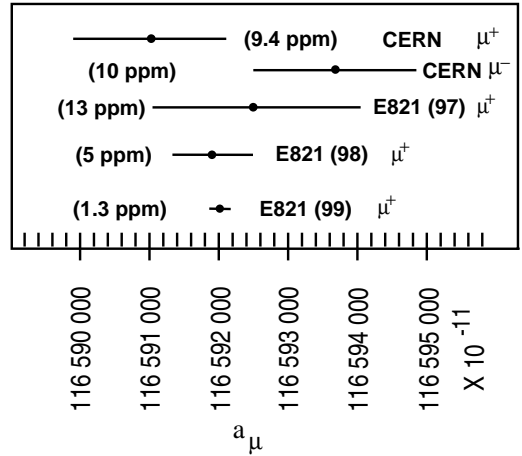}
\caption{Experimental measurements of $a_{\mu}$.}
\label{fig:answer}
\end{figure}

The theoretical and experimental values for the $a_{\mu}$
are not known nearly as well as for $a_e$.
In 1999, prior to the new published result,
the world average of measured values was
$a_{\mu}^{exp}=(1165920.5\pm 4.6)\times 10^{-9}$ (4 ppm).
This included 9.4 and 10 ppm results for the positive and negative
muon, respectively, in a series
of famous experiments at CERN ending in the 1970's,\cite{CERN}
as well as the results
from the new Brookhaven muon (g-2) experiment (E821)
using positive muon data taken in 1997 (13 ppm)\cite{E821a} and 1998
(5 ppm)\cite{E821b}. All of these measurements are in agreement within their
errors, as shown in Fig.~\ref{fig:answer}; the negative muon data were
incorporated under the assumption of CPT invariance.
Using the recent published compilation of the theoretical
ingredients to $a_{\mu}^{th}$ by
Czarnecki and Marciano,\cite{CM}
$a_{\mu}^{th}=(1165915.96\pm 0.67)\times 10^{-9}$ (0.6 ppm),
which used the hadronic evaluation by Davier and H\"ocker,\cite{DH98b}
we find that agreement between experiment and data was good:
$a_{\mu}^{exp}-a_{\mu}^{th}=(4.5\pm 4.7)\times 10^{-9}$.
Recently, an error in the hadronic light-by-light contribution was
found,\cite{KN}
which changes the theoretical prediction to 
$a_{\mu}^{th}=(1165917.68\pm 0.67)\times 10^{-9}$ (0.6 ppm).
This improves the agreement between theory and experiment:
$a_{\mu}^{exp}-a_{\mu}^{th}=(2.8\pm 4.7)\times 10^{-9}$.

The new experimental result for $a_{\mu^+}$,
based on a sample of 1 billion positrons
collected in 1999 by E821,\cite{E821c} is 
$a_{\mu}^{exp}(E821)=(1165920.2\pm 1.4 \pm 0.6)\times 10^{-9}$ (1.3 ppm).
The new world value changes very little,
$a_{\mu}^{exp}=(1165920.3\pm 1.5)\times 10^{-9}$ (1.3 ppm).
Comparing with the uncorrected theoretical number
gives a 2.6 $\sigma$ difference between measurement and theory:
$\Delta a_{\mu}=a_{\mu}^{exp}-a_{\mu}^{th}=(4.3\pm 1.6)\times 10^{-9}$
($3.7 \pm 1.4$ ppm). Including the light-by-light correction,
$\Delta a_{\mu}=a_{\mu}^{exp}-a_{\mu}^{th}=(2.6\pm 1.6)\times 10^{-9}$
($2.2 \pm 1.4$ ppm), a $1.6\sigma$ difference.

Two additional data sets, from 2000 and 2001 runs, are currently
being analyzed. The 2000 data set consists of about 4 billion events
for the $\mu^+$ and  2001 consists of about 3 billion events for the
$\mu^-$. The original stated goal of E821 was
to reduce the anomaly measurement error to 0.35 ppm. Ideally, one
would also equalize the
errors on the $\mu^+$ and $\mu^-$ in order to optimally test CPT invariance and
to study systematic issues in the experiment.
We expect to come close to this goal, but this
will require a future data run with 6 billion events.

\section{Theory Status}

The Standard Model contributions to $a_{\mu}$
can be conveniently separated into QED, electroweak, and hadronic
portions. Although the anomaly is dominated by the QED contribution,
there are significant hadronic and EW contributions at the level
of $\approx 58.3 $ ppm
and $\approx 1.3$ ppm, respectively. 
We discuss each of these contributions below, with
emphasis on the hadronic contribution,
whose error dominates the overall error in $a_{\mu}^{th}$.

The QED contribution, using $\alpha$ from $a_e^{exp}$ and calculated to fifth
order in  ${\alpha \over \pi}$ (some of the highest order diagrams
were estimated), contributes\cite{CM}
an error of 21 ppb to $a_{\mu}^{th}$,
$a_{\mu}^{QED}=116584705.7(2.9) \times 10^{-11}$.
Kinoshita\cite{TK} reports that some of the
fourth order terms are being re-calculated with 128 bit precision,
which may result in a small shift in the QED contribution; this is
expected to have minimal impact on the comparison between experimental and
theoretical values of $a_{\mu}$.

\begin{figure}
\epsfxsize120pt
\includegraphics[angle=0,width=.48\textwidth]
{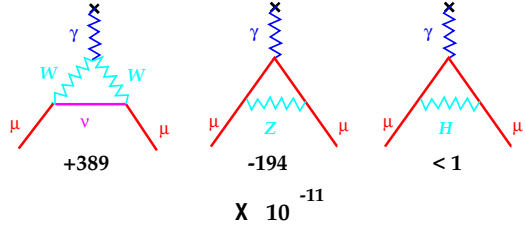}
\caption{Lowest-order electroweak contributions.}
\label{fig:EW}
\end{figure}

The lowest order electroweak diagrams, involving the exchange of
a W, Z or Higgs, are shown in Fig.~\ref{fig:EW}.
The electroweak contribution, including the
25\% reduction from higher order terms, is\cite{CM}
$a_{\mu}^{EW}=151(4) \times 10^{-11} (1.30 \pm 0.03)$ppm.
The theoretical uncertainty is very small. The central value
is right on the edge of the current experimental error,
and it will be a significant
contribution at the experimental error goal of
0.35 ppm.

\begin{figure}
\epsfxsize120pt
\includegraphics[angle=0,width=.48\textwidth]
{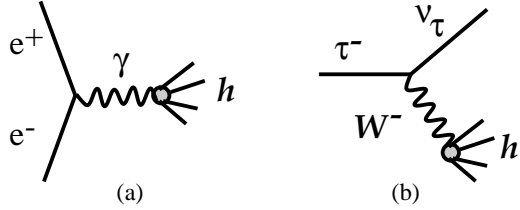}
\caption{The electron scattering and tau decay diagrams relevant to the
determination of the lowest order contribution to $a_{\mu}$.}
\label{fig:hadpro}
\end{figure}

The lowest order hadronic diagram is shown in
Fig.~\ref{fig:had1}, where a hadron loop
has been inserted into the Schwinger diagram.
Since these contributions involve
the strong interaction at low energies, they cannot be calculated from first
principles.
Their contribution
can, however, be determined from measured $e^+e^-$ scattering
cross sections over all energies through the use of the dispersion
relation, Eq. ~\ref{eq:disp}, (also see Fig.~\ref{fig:hadpro}(a))

\begin{equation}
a_{\mu}(had; 1)=({\alpha m_{\mu} \over 3\pi})^2\int_{4m_{\pi}^2}^{\infty}
{ds  \over s^2}K(s)R(s)
\label{eq:disp}
\end{equation}
$K(s)$ is a slowly varying function and the $e^+e^-$ data enter through
the ratio of cross sections,
$R(s)=
{\sigma(e^+e^-\rightarrow hadrons)
\over\sigma(e^+e^-\rightarrow \mu^+\mu^-)}$.
The low energy data are the most important as a result of
their large amplitude and the ${1 \over s^2}$ term in the integrand, where
$\sqrt{s}$ is the center-of-mass energy.

High quality hadronic $\tau$ decay data from LEP and Cornell
can be used to augment the isovector part of the $e^+e^-$ data 
at energies below $m_{\tau}c^2$
(second diagram in Fig.~\ref{fig:hadpro})
using isospin invariance (to relate for example $\pi^- \pi^0$ channels in
$\tau$ decays
to
 $\pi^+ \pi^-$ channels in $e^+e^-$ collisions) and the CVC hypothesis
(to connect the $W^-$ and photon intermediate states).\cite{ADH98}

There have been a number of
evaluations of $a_{\mu}$(had;1) over the past two decades.
We will look in some detail at those performed since 1995.
The contributions of $e^+e^-$ data to $a_{\mu}(had;1)$
in different energy ranges,
from the evaluation by Brown and Worstell,\cite{BW96}
are indicated in Table~\ref{tab:BWhad}. (We note that the new data
which are becoming available will render this table obsolete.)
The largest contribution to the value and error are for
$\sqrt{s}<1.4 $ GeV, a region dominated by the effects of the $\rho$ resonance
(see Figs. 6-8).
The other energy ranges give a considerably smaller contribution to the
error, however
for an $0.35 $ ppm measurement they  cannot be neglected, particularly in the
range $1.4 \rightarrow 2.6 $ GeV.

\begin{table}
\caption{As of 1995, contributions to $a_{\mu}$(had;1) as a function
of $e^+e^-$
energy, illustrating the (soon-to-be obsolete)
sources of error.\protect\cite{BW96}. With the
incorporation of new $e^+e^-$ and $\tau$ data, the errors below 5 GeV
decrease significantly from the values shown in the table.}
\label{tab:BWhad}
\begin{tabular}{|c|c|c|} 
 
\hline 
 
\raisebox{0pt}[12pt][6pt]{$\sqrt{s}, $GeV} & 
 
\raisebox{0pt}[12pt][6pt]{$a_{\mu}(had;1)$} & 
 
\raisebox{0pt}[12pt][6pt]{Error, ppm}\\
 
\hline
 
\raisebox{0pt}[12pt][6pt]{$< 1.4$} & 
 
\raisebox{0pt}[12pt][6pt]{87. \%} & 
 
\raisebox{0pt}[12pt][6pt]{1.29}\\
\cline{1-3}
\raisebox{0pt}[12pt][6pt]{$1.4 \rightarrow 2.0$} & 
 
\raisebox{0pt}[12pt][6pt]{4.6 \%} & 
 
\raisebox{0pt}[12pt][6pt]{0.21}\\

\cline{1-3}
\raisebox{0pt}[12pt][6pt]{$2.0 \rightarrow 3.1$} & 
 
\raisebox{0pt}[12pt][6pt]{4.0 \%} & 
 
\raisebox{0pt}[12pt][6pt]{0.30}\\

\cline{1-3}
\raisebox{0pt}[12pt][6pt]{\ \ \ $2.0 \rightarrow 2.6$} & 
 
\raisebox{0pt}[12pt][6pt]{2.9 \%} & 
 
\raisebox{0pt}[12pt][6pt]{0.27}\\

\cline{1-3}
\raisebox{0pt}[12pt][6pt]{\ \ \ $2.6 \rightarrow 3.1$} & 
 
\raisebox{0pt}[12pt][6pt]{1.1 \%} & 
 
\raisebox{0pt}[12pt][6pt]{0.12}\\

\cline{1-3}
\raisebox{0pt}[12pt][6pt]{$J/\Psi$ (6 states)} & 
 
\raisebox{0pt}[12pt][6pt]{1.3 \%} & 
 
\raisebox{0pt}[12pt][6pt]{0.08}\\

\cline{1-3}
\raisebox{0pt}[12pt][6pt]{QCD $3.1 \rightarrow \infty$} & 
 
\raisebox{0pt}[12pt][6pt]{3.0 \%} & 
 
\raisebox{0pt}[12pt][6pt]{0.03}\\

\cline{1-3}
\raisebox{0pt}[12pt][6pt]{Total} & 
 
\raisebox{0pt}[12pt][6pt]{} & 
 
\raisebox{0pt}[12pt][6pt]{1.37}

\\\hline
\end{tabular}
\end{table}

Since 1995, there has been a substantial improvement in the data quality,
but major portions of the data are either preliminary or are
still in the process
of being incorporated into evaluations of  $a_{\mu}$(had;1).

\begin{figure}
\epsfxsize120pt
\includegraphics[angle=0,width=.5\textwidth]
{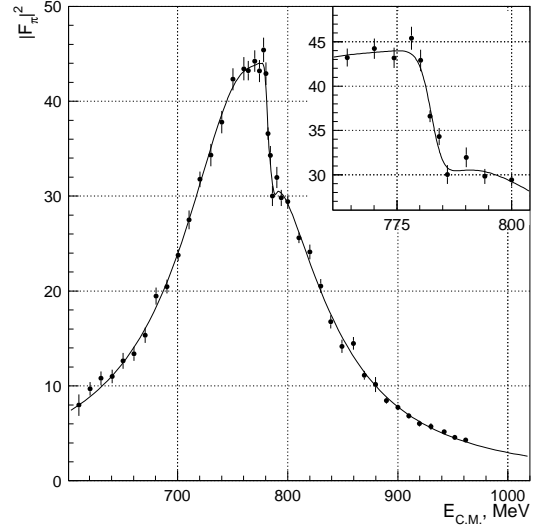}
\caption{Preliminary pion form factor data from Novosibirsk in the vicinity
of the $\rho$ resonance.\protect\cite{VEPP2M} Inset: data the $\omega$
interference region.}
\label{fig:VEPP2M}
\end{figure}

Data on the pion form factor (which can be directly related to the
$e^+e^-$ cross section) from the CMD2 and SND 
experiments at the VEPP-2M accelerator in Novosibirsk are nearing publication.
They cover the important energy range
$\sqrt{s}<1.4 $ GeV. Preliminary
data in the $\rho$ resonance range (600 to 930 MeV)
from CMD2 are shown in
Fig.~\ref{fig:VEPP2M}.\cite{VEPP2M}
Their anticipated systematic
error of 0.6\% in this range
would reduce the 
error contribution from the $e^+e^-$ data in this energy
region by better than
a factor of two. The VEPP-2000 project is an upgrade under
construction at Novosibirsk which
will extend quality $e^+e^-$ measurements up to 2 GeV,
with an order of magnitude or more
improvement in luminosity.

\begin{figure}
\epsfxsize120pt
\includegraphics[angle=0,width=.5\textwidth]
{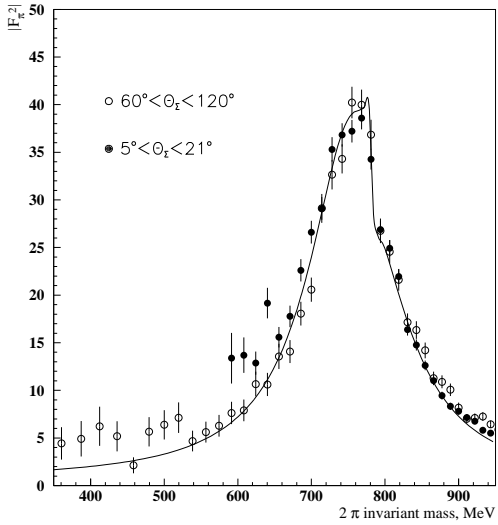}
\caption{Preliminary pion form factor data from
radiative return measurements (KLOE).\protect\cite{KLOE}}
\label{fig:KLOE}
\end{figure}

Data relevant to the low energy region are also being taken
by the KLOE\cite{KLOE} experiment at DA$\Phi$NE. They operate at the
$\phi$ mass, and then derive $e^+e^-$ cross sections at lower energies
using the so-called radiative return method. In this approach, the
initial state  electron or positron radiates away some of its energy
via a photon, providing access to scattering at energies below the
$\phi$ mass. Preliminary data are shown in Fig.~\ref{fig:KLOE}.
Currently their systematic errors are on the order of a few percent, however,
they expect to improve this in the near future to the point that
they are competitive with the Novosibirsk data.
This measurement, along with possible plans at other $e^+e^-$
machines such as the B-factories and Cornell
which have access to higher energies,
will contribute significantly to the next wave of precision $e^+e^-$ data.

\begin{figure}
\epsfxsize120pt
\includegraphics[angle=-90,width=.5\textwidth] {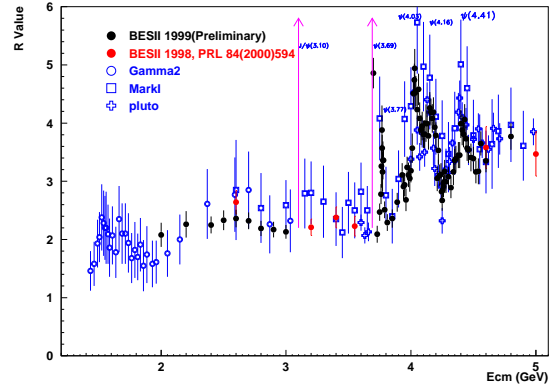}
\caption{Preliminary and published $e^+e^-$ data, 2-5 GeV,
from BESII,
along with older data from other experiments\protect\cite{BES}.}
\label{fig:BES}
\end{figure}

There are new, preliminary and published data from BESII\cite{BES}
in the important
energy range $2-5 $ GeV (Fig.~\ref{fig:BES}). Note especially that in the range
$2-3 $ GeV, which has the largest contribution to $a_{\mu}$(had;1),
BESII data have much smaller errors and the central values
are 15\% lower compared to the old MARK I and Gamma2 data.

High quality $\tau$ decay data, which have been produced
over the last few years as described by
P. Roudeau at this conference,  have significantly
reduced the contribution to the error of $a_{\mu}$(had;1) for energies
below the $\tau$ mass.\cite{ADH98} In addition,
new analysis results should be available soon from ALEPH.\cite{davier}
Some authors\cite{eidel}$^,$\cite{melnikov} have questioned, however,
whether there are sufficient controls over radiative corrections
and corrections to the approximations of CVC and isospin invariances.
In addition, there appears to be an overall normalization
disagreement between the $\tau$
data from ALEPH and Cornell.\cite{urheim}
The $\tau$ data represent, nevertheless,
valuable additions and checks on the $e^+e^-$ data at low energies,
and various groups plan to study these issues in the future.

A list of $a_{\mu}$(had;1) evaluations is given in
Table~\ref{tab:had1}. Various combinations of ingredients were used:
$e^+e^-$ data from a wide range of experiments,
$\tau$ decay data from LEP and CESR,
or theoretical input from perturbative QCD (pQCD)
in the higher energy regions where
it can be relied upon but where the data quality is poor.

Eidelman and Jegerlehner\cite{EJ95}, EJ95, relied primarily on
$e^+e^-$ data, using pQCD only at the highest energies ($>20 $ GeV).
Brown and Worstell\cite{BW96},
BW96, used essentially the same data set as EJ95, but took into account
the correlation of errors among data points coming from the same experiment.
BW96 and EJ95 values and errors are in excellent agreement,
suggesting that the correlation
issue was not so important.
Adel and Yndurain\cite{AY95}, AY95, used pQCD in regions where the $e^+e^-$
data were poor, and obtained a value somewhat higher than, but still in
agreement with, EJ95 and BW96. The theoretical input enabled hem to quote
a smaller error.
In 1998, Adelman, Davier
and H\"ocker,\cite{ADH98} ADH98,
did a full re-evaluation using updated values of the data sets used
by EJ95 and BW96, obtaining the same error and a central value which
was lower but still consistent within errors.
ADH98 then incorporated the high quality $\tau$ data from ALEPH,
producing a dramatic improvement in the error of the contribution
below the $\tau$ mass, and reducing the overall error in $a_{\mu}$
by 40\%; inclusion of $\tau$ data increased
the central value somewhat.
Subsequently, Davier and H\"ocker\cite{DH98a}, DH98a, refined the
$e^+e^-$ and $\tau$ evaluation of ADH98 by applying pQCD 
above 1.8 GeV in regions where the data were poor,
resulting in another 20\% decrease in the error 
and a decrease in the central value. Most of the changes
can be attributed to the
effect of using pQCD from 1.8 to 3 GeV, where the old data are rather poor,
as seen in Fig.~\ref{fig:BES}.
The DH98a evaluation
was done before the new BESII data (Fig.~\ref{fig:BES})
were available in this energy range;
their pQCD calculations are in very good agreement with the BESII data,
and about 15\% below the old data, pointing to the reliability of pQCD at
these energies.
In a subsequent work (DH98b) the same authors
applied QCD sum rules at low energies, resulting in slight further reductions
in error and central value. The value from DH98b was used in the
Czarnecki and Marciano\cite{CM} theoretical compilation and also by
E821\cite{E821c} to compare with their new experimental number.

\begin{table}
\caption{A list of a number of recent evaluations of $a_{\mu}(had;1)$.
The last entry is the E821 experimental 'measurement' of $a_{\mu}$(had;1),
obtained by subtracting the QED, EW and higher order hadronic contributions
from the experimental number. This is an updated version of a similar table
found in Ref.\protect\cite{MR}.}
\label{tab:had1}
\begin{tabular}{|c|c|c|} 
 
\hline 
 
\raisebox{0pt}[12pt][6pt]{Ref.} & 
 
\raisebox{0pt}[12pt][6pt]{$a_{\mu}(had;1)$} & 
 
\raisebox{0pt}[12pt][6pt]{Comment}\vspace{-.2cm}\\

\raisebox{0pt}[12pt][6pt]{} & 

\raisebox{0pt}[12pt][6pt]{($\times 10^{11}$)} & 

\raisebox{0pt}[12pt][6pt]{}\\

\hline
 
\raisebox{0pt}[12pt][6pt]{EJ95\cite{EJ95}} & 
 
\raisebox{0pt}[12pt][6pt]{7024(153)} & 
 
\raisebox{0pt}[12pt][6pt]{$e^+e^-$}\\
\cline{1-3}
\raisebox{0pt}[12pt][6pt]{BW96\cite{BW96}} & 
\raisebox{0pt}[12pt][6pt]{7026(160)} & 
 
\raisebox{0pt}[12pt][6pt]{$e^+e^-$}\\

\cline{1-3}
\raisebox{0pt}[12pt][6pt]{AY95\cite{AY95}} & 
 
\raisebox{0pt}[12pt][6pt]{7113(103)} & 
 
\raisebox{0pt}[12pt][6pt]{$e^+e^-$,QCD}\\

\cline{1-3}
\raisebox{0pt}[12pt][6pt]{ADH98\cite{ADH98}} & 
 
\raisebox{0pt}[12pt][6pt]{6950(150)} & 
 
\raisebox{0pt}[12pt][6pt]{$e^+e^-$}\\

\cline{1-3}
\raisebox{0pt}[12pt][6pt]{ADH98\cite{ADH98}} & 
 
\raisebox{0pt}[12pt][6pt]{7011(94)} & 
 
\raisebox{0pt}[12pt][6pt]{$e^+e^-, \tau$}\\

\cline{1-3}
\raisebox{0pt}[12pt][6pt]{DH98a\cite{DH98a}} & 
 
\raisebox{0pt}[12pt][6pt]{6951(75)} & 
 
\raisebox{0pt}[12pt][6pt]{$e^+e^-, \tau$}\vspace{-.2cm}\\

\raisebox{0pt}[12pt][6pt]{} & 
 
\raisebox{0pt}[12pt][6pt]{} & 
 
\raisebox{0pt}[12pt][6pt]{pQCD}\\

\cline{1-3}
\raisebox{0pt}[12pt][6pt]{DH98b\cite{DH98b}} & 
 
\raisebox{0pt}[12pt][6pt]{6924(62)} & 
 
\raisebox{0pt}[12pt][6pt]{$e^+e^-, \tau$}
\vspace{-.2cm}\\

\raisebox{0pt}[12pt][6pt]{} & 
 
\raisebox{0pt}[12pt][6pt]{} & 
 
\raisebox{0pt}[12pt][6pt]{pQCD, sum rules}\\

\cline{1-3}
\raisebox{0pt}[12pt][6pt]{N01\cite{N01}} & 
 
\raisebox{0pt}[12pt][6pt]{7021(76)} & 
 
\raisebox{0pt}[12pt][6pt]{$e^+e^-, \tau$}\\

\cline{1-3}
\raisebox{0pt}[12pt][6pt]{TY01\cite{TY01}} & 
 
\raisebox{0pt}[12pt][6pt]{6966(73)} & 
 
\raisebox{0pt}[12pt][6pt]{$e^+e^-, \tau$}\vspace{-.2cm}\\

\raisebox{0pt}[12pt][6pt]{} & 
 
\raisebox{0pt}[12pt][6pt]{} & 

\raisebox{0pt}[12pt][6pt]{space-like $F_{\pi}$}\\

\cline{1-3}
\raisebox{0pt}[12pt][6pt]{E01\cite{E01}} & 
 
\raisebox{0pt}[12pt][6pt]{6932(65)} & 
 
\raisebox{0pt}[12pt][6pt]{$e^+e^-$}\\

\cline{1-3}
\raisebox{0pt}[12pt][6pt]{E821\cite{E821c}} & 
 
\raisebox{0pt}[12pt][6pt]{7350(153)} & 
 
\raisebox{0pt}[12pt][6pt]{Expt -[ QED}\vspace{-.2cm}\\

\raisebox{0pt}[12pt][6pt]{} & 
 
\raisebox{0pt}[12pt][6pt]{} & 
 
\raisebox{0pt}[12pt][6pt]{+EW+(Had$>$1)]}

\\\hline
\end{tabular}
\end{table}

Some preliminary and published theoretical evaluations
of $a_{\mu}$(had;1) have appeared since the E821 publication
of Brown, et al.\cite{E821c} Narison\cite{N01}, N01, has used essentially
the same $\tau$ and $e^+e^-$ data sets at low energies and for the
resonances as ADH98,
with QCD applied to the continuum at the higher energies ($>$ 1.7 GeV),
arriving at nearly the same value as ADH98, but with a slightly smaller error.
Troconiz and Yndurain\cite{TY01}, TY01,
have applied the maximum available data (including some preliminary data)
and theory,
following the earlier approach of AY95, and also incorporated
pion form factor data from pion scattering at low energies,
to arrive at a value
which compares closely to
DH98b, with a slightly larger central value and error.

Finally, Eidelman\cite{VEPP2M}, E01,
has produced a preliminary number based entirely on
$e^+e^-$ data (except for pQCD at the very highest energies),
including the new preliminary results from Novosibirsk and
BESII.
He obtains a value and error which are almost the same as DH98b.

What can we conclude from this series of $a_{\mu}(had;1)$ determinations?
The first thing we note is
that
within their stated errors,
all of the evaluations are in agreement.
In particular, the most recent evaluations are in excellent agreement
even with their smaller errors, although we note that the DH98b evaluation
has the smallest value leading to the largest discrepancy between theory
and experiment.
Secondly, an analysis which incorporates all of the new $\tau$ and
$e^+e^-$ data would be helpful.
Eidelman, Davier and H\"ocker are presently collaborating
on such an evaluation.
When the final analyses, which include the latest excellent data
are completed, we can expect a more reliable value for $a_{\mu}(had;1)$ with a
smaller error.

\begin{figure}
\epsfxsize120pt
\includegraphics[angle=0,width=.5\textwidth] {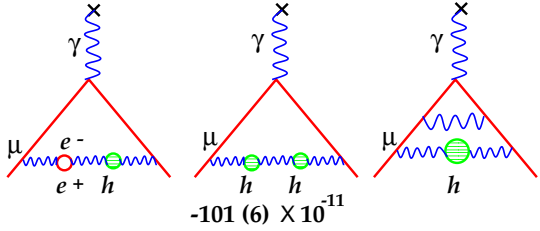}
\caption{Some diagrams of the higher-order hadronic contributions.}
\label{fig:had2}
\end{figure}

\begin{figure}
\epsfxsize120pt
\includegraphics[angle=0,width=.25\textwidth] {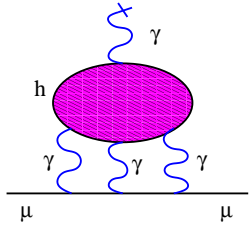}
\caption{
Diagram for the hadronic light-on-light contribution.}
\label{fig:lol}
\end{figure}

The higher order hadronic contributions, $a_{\mu}(had; >1)$,
can be separated into two parts.
One part, involving
higher order diagrams such as those in Fig.~\ref{fig:had2}, have a
relatively small contribution to $a_{\mu}$ and the error is
negligible:\cite{KR}
$a_{\mu}(had;>1)=-101(6)\times 10^{-11}$.
The other part, involving the hadronic light-on-light diagram in
Fig.~\ref{fig:lol} (LOL), presents
special problems because, unlike other hadronic terms,
it cannot be estimated based on experimental data.
The value used in Czarnecki and Marciano\cite{CM}
is an average of the values and the
errors of two separate determinations\cite{HK}$^,$ \cite{BPP}
(which are in agreement within errors),
$a_{\mu}$(had;LOL)$=-85(25)\times 10^{-11}$.
Both calculations use models motivated by chiral
perturbation theory to calculate the contributions at low
energies.
The largest contribution comes from
the $\pi^0$ pole term, with lesser contributions
from the $\eta$ and $\eta'$ poles. Other contributions
can be $\approx 20-30$\% in size relative to the pole terms, but when added
together they largely cancel.
After the publication of the new E821 experimental result\cite{E821c}
and after this
talk was given at LP01, Knecht and Nyffeler
\cite{KN} calculated the pion pole
contribution using Large-$N_C$ and short-distance properties of QCD.
They obtained virtually the same magnitude as in
references \cite{HK} and \cite{BPP} but with the opposite sign:
$a_{\mu}^(had;LOL,\pi^0)=+5.8(1.0)\times 10^{-10}$.
The signs of the $\eta$ and $\eta'$ poles also change, however,
Knecht and Nyffeler only estimated their magnitude using a VMD model;
their total pseudo-scalar contribution is
$a_{\mu}(had;LOL,PS)=+8.3(1.2)\times 10^{-10}$.
The sign error has since
been acknowledged by the authors of \cite{HK} and \cite{BPP} (see references
\cite{HKa} and \cite{BPPa}). 
Changing the signs of the pole and axial vector terms,
but keeping the other terms
as calculated in \cite{HK} and \cite{BPP} the same, increases the
theoretical value of $a_{\mu}$ by $17.2 \times 10^{-10}$, very close to just
reversing the sign on the entire LOL contribution.
Knecht and Nyffeler plan in the future to calculate all of the other terms
in the hadronic LOL contribution.
Several other groups are also considering new ways to tackle
this difficult calculation.
Ultimately, the hadronic LOL term may prove to be the
limiting factor in the theoretical error of $a_{\mu}$.

The theoretical error on $a_{\mu}(had;1)$
has gone down dramatically over
the years, and with new, more accurate calculations
of the light-by-light contribution, the error and reliability 
of $a_{\mu}^{th}$ should continue to improve significantly.

\begin{figure}
\epsfxsize120pt
\includegraphics[angle=0,width=.48\textwidth] {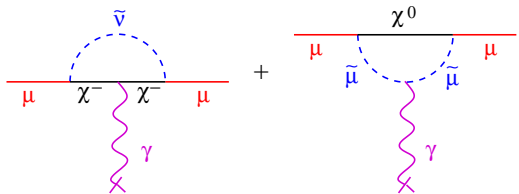}
\caption{
Lowest order diagrams for SUSY contributions.}
\label{fig:SUSY}
\end{figure}

New physics will be reflected by a non-zero value of
$\Delta a_{\mu}^{new}=a_{\mu}^{exp}-a_{\mu}^{th}$.
Some examples are muon substructure, anomalous gauge couplings,
leptoquarks, or supersymmetry.
In a minimal supersymmetric model with degenerate sparticle
masses (see Fig.~\ref{fig:SUSY}),
the contribution to $\Delta a_{\mu}^{new}$ would be substantial in the
case where $\tan{\beta}$ is large:
$\Delta a_{\mu}^{SUSY}\approx
140 \times 10^{-11}({100 GeV \over \tilde m})^2\tan{\beta}$, where
$\tilde m$ is the sparticle mass.
For $4< \tan{\beta}<40$, $\tilde m \approx 150-500$ GeV.
Note that the supersymmetric diagrams (Fig~\ref{fig:SUSY})
are analogous to the electroweak diagrams (Fig~\ref{fig:EW}).
If supersymmetry is to explain all of
$\Delta a_{\mu}^{new}$, then its contribution
is large: almost two  times bigger than the electroweak contribution.
Of course, the presently observed $\Delta a_{\mu}$ may also be due to
a statistical variation in the experimental number
or to errors in the experiment or theory.
When the additional data are analyzed, and with the continued
extensive studies of the theory,  one can anticipate
that major progress will be made in understanding  any non-zero
$\Delta a_{\mu}$.

\section{Experiment}

The ongoing muon (g-2) experiment at Brookhaven National Laboratory,
E821, had its beginnings in the early 1980's. Its original goal
was to measure
$a_{\mu}$ to 0.35 ppm, or about 20 times better than the CERN\cite{bailey}
experiment.
It received Laboratory
approval in 1987,
and major construction on the storage ring magnet began in the early 1990's.
The first data were taken in 1997, with one major run in each year
1998-2001.
The experimental technique follows the general one used in the CERN experiment
with a number of important improvements and innovations.

The Brookhaven AGS
delivers up to $7\times 10^{12}$ protons per bunch, with
energies of 24 GeV, onto a water-cooled,
rotating nickel target.
There are 6-12 bunches per $\approx 2.5$ second AGS cycle,
each about 50 ns wide FWHM and spaced 33 milliseconds apart.
Secondary pions emitted from the target with momenta of 3.1 GeV/$c$
are sent down a 72 m straight section
of alternating magnetic quadrupoles, where highly polarized muons
from forward pion decays are collected.
The beam is then momentum-selected for either pions or the slightly
lower-momentum muons, and then is
injected through a field-free inflector\cite{inflector}
region into a circular storage ring possessing
a very homogeneous magnetic field.

For the case of pion injection,
with the pion momentum slightly higher than that of the central storage
ring momentum,
a small fraction ($\approx 25 $ ppm)
of the muons from pion decays will have the correct momenta and directions to
be stored.
The efficiency of this process is low, and the very high intensity of
pions and secondary particles
associated with pion interactions with surrounding materials
creates severe background problems for the detector
system near the injection time ('flash'). An essential improvement
over prior experiments was the incorporation of direct muon injection.
With muon injection, the number of stored muons
is increased by a factor of 10, while the 'flash' is reduced by a factor of
50 because most of the higher-momentum pions are blocked by beam-line
collimators. In the homogeneous B field of the storage ring, charged
particles follow a circular path (slightly modified by the
electric quadrupole field) which would cause
them to strike the inflector after one revolution.
Muon injection therefore 
requires an in-aperture magnetic pulse at $90^0$ around the ring from
the injection point (provided by the pulsed 'kicker')
in order to center the muon orbits in the storage region.

With either muon or pion injection,
positive (negative) muons are stored in the ring
with spins initially polarized anti-parallel
(parallel) to their momenta. In a magnetic field (no E-field)
the spins precess relative to the muon momenta according to
$\vec \omega_a=\vec \omega_s-  \vec \omega_c =-a_{\mu}{e\vec B \over mc}$.
Here $\omega_a$ and $\omega_c$ are the angular frequencies
of spin rotation and momentum rotation (or cyclotron angular frequency),
respectively.
Note that all of the muons precess at the same rate in a given field,
regardless of their momenta.
Two quantities must be measured with precision to determine $a_{\mu}$:
$\omega_a$ and $B$, each being
time averaged over the ensemble of muons.
Actually, instead of measuring B,
we determine the frequency of precession of the free proton, $\omega_p$,
in the same average magnetic field as the muons via NMR measurements.
The anomaly is given by Eq. ~\ref{eq:geta},

\begin{equation}
a_{\mu}={R \over \lambda -R}, 
\label{eq:geta}
\end{equation}

\noindent
where $R={<\omega_a> \over <\omega_p>}$.
$\lambda={\mu_{\mu} \over \mu_p}=3.183\ 345\ 39(10)$
is the ratio of the muon and proton
magnetic moments determined from other
experiments\cite{liu}$^,$\cite{pdg2000}.
The analyses of $<\omega_a>$ and $<\omega_p>$ were independent, and furthermore
concealed offsets were maintained in each value so that no one could
calculate $a_{\mu}$ prior to the completion of the analyses.

The average trajectories of muons in the storage ring can only be
known moderately well.
Therefore the B field needs to be as uniform as possible to minimize the
dependence of a given muon's precession rate on its exact trajectory in the
storage ring. This prohibits the use of a gradient magnetic field
to store (focus) the beam.
E821 follows the CERN approach of using a quadrupole electrostatic field to
provide the focusing.
In the presence of the electric field,
the precession is described by Eq.~\ref{eq:modomega},

\begin{equation}
\vec \omega_a=-{e \over mc}[a_{\mu}\vec B - (a_{\mu} - {1 \over \gamma^2 -1})
\vec \beta \times \vec E]
\label{eq:modomega}
\end{equation}

The ``magic'' $\gamma\approx 29.3$, or $p_{\mu}\approx 3.094$ GeV/$c$,
is chosen so that
$a_{\mu} - {1 \over \gamma^2 -1}\approx 0$, minimizing the effect of $\vec E$ on
$\vec \omega_a$.
Because not all stored muons have the
exact magic momentum, a small correction ($\approx 0.6 $ ppm)
must be applied to the final
value for $\omega_a$. Equation 3 is strictly valid only for muon motion
perpendicular to B; the up-down motion associated with vertical betatron
oscillations leads to another small correction of
$\approx 0.2 $ ppm, the so-called ``pitch correction''.

\begin{figure}
\epsfxsize120pt
\includegraphics[angle=0,width=.5\textwidth]
{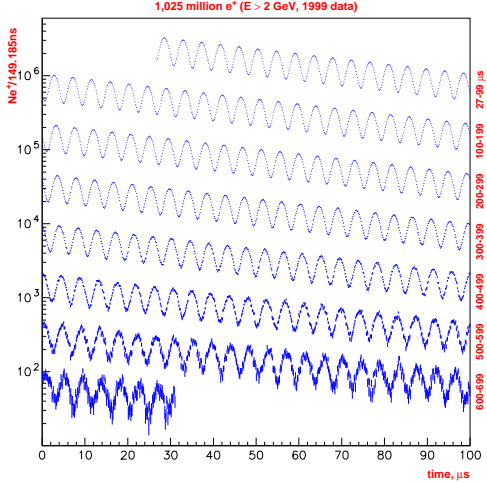}
\caption{Spectrum of number of positrons versus time,
from the 1999 data sample. There are a total of 1 billion $e^+$
above 2 GeV.}
\label{fig:data}
\end{figure}

The practical limit to the strength of a ferric field with the
required homogeneity is $\approx 1.5$T;
E821 chose $B=1.45 $T, leading to a ring
radius of $7.112 $m.  The storage ring aperture radius
is 4.5 cm, giving a $\approx\pm 0.6\%$ ($\approx\pm 0.4\%$) base-to-base
range in stored momenta for pion (muon) injection.
The cyclotron period is
$\tau_c={1\over f_c}={2 \pi\over \omega_c}\approx 149.2 $ns, the
precession period is $\tau_a\approx 4.365\ \mu$s and the dilated muon lifetime
is $\tau = \gamma \tau_0 \approx 64.38\ \mu $s. Decays are typically measured
for at least ten muon lifetimes, or about 4000 cyclotron and
150 precession periods. A log plot of the 1999 data set, folded into
$100 \mu s$ periods, is shown
in Fig.~\ref{fig:data}.

The error on $a_{\mu}$ from the combined $\mu^+$ and $\mu^-$ data sets
from the CERN\cite{CERN} g-2 experiment
is 7 ppm with a 1.5 ppm systematic error. By comparison,
E821 must keep the systematic errors in B($\omega_p$) and $\omega_a$ to
less than a few tenths
of a ppm to approach its experimental precision goal of 0.35 ppm.

\subsection{The Magnet and the Determination of $\omega_p$}

\begin{figure}
\epsfxsize120pt
\includegraphics[angle=0,width=.5\textwidth]
{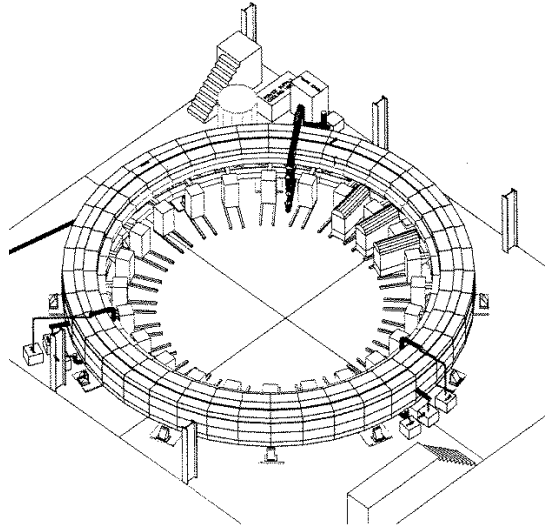}
\caption{Overhead schematic view of the storage ring magnet. The detectors
are distributed in uniform intervals around the inside of the ring.
The beam is brought in through a hole in the back of the magnet yoke, indicated
by the solid line at 10 o'clock.}
\label{fig:ring_view}
\end{figure}

\begin{figure}
\epsfxsize120pt
\includegraphics[angle=0,width=.5\textwidth]
{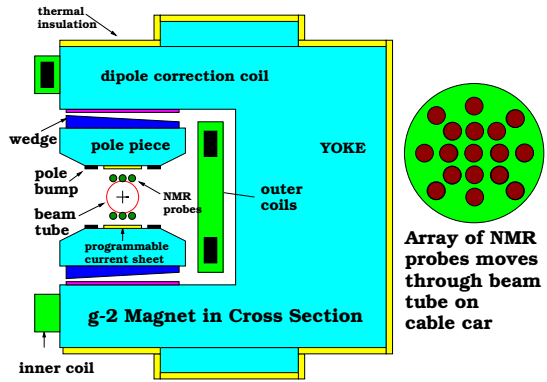}
\caption{Cross section view of the storage ring magnet. The
means of shimming the magnet and measuring its field are indicated.}
\label{fig:magnetland}
\end{figure}

The storage ring,\cite{danby} Fig.~\ref{fig:ring_view},
is a continuous C-magnet open to the inside. A cross-section view,
Fig.~\ref{fig:magnetland}, shows its essential features.
It contains more than 600  tons of magnet steel.
Three superconducting coils, which provide exceptional
B-field stability with time, are used to power the magnet.
The entire magnet is wrapped in thermal insulation to reduce gap
changes due to temperature change.
The storage region of 4.5 cm radius is defined by a series of circular
collimators inside an evacuated chamber.
The pole gap is 18 cm high and 53 cm wide.

\begin{figure}
\epsfxsize120pt
\includegraphics[angle=0,width=.5\textwidth]
{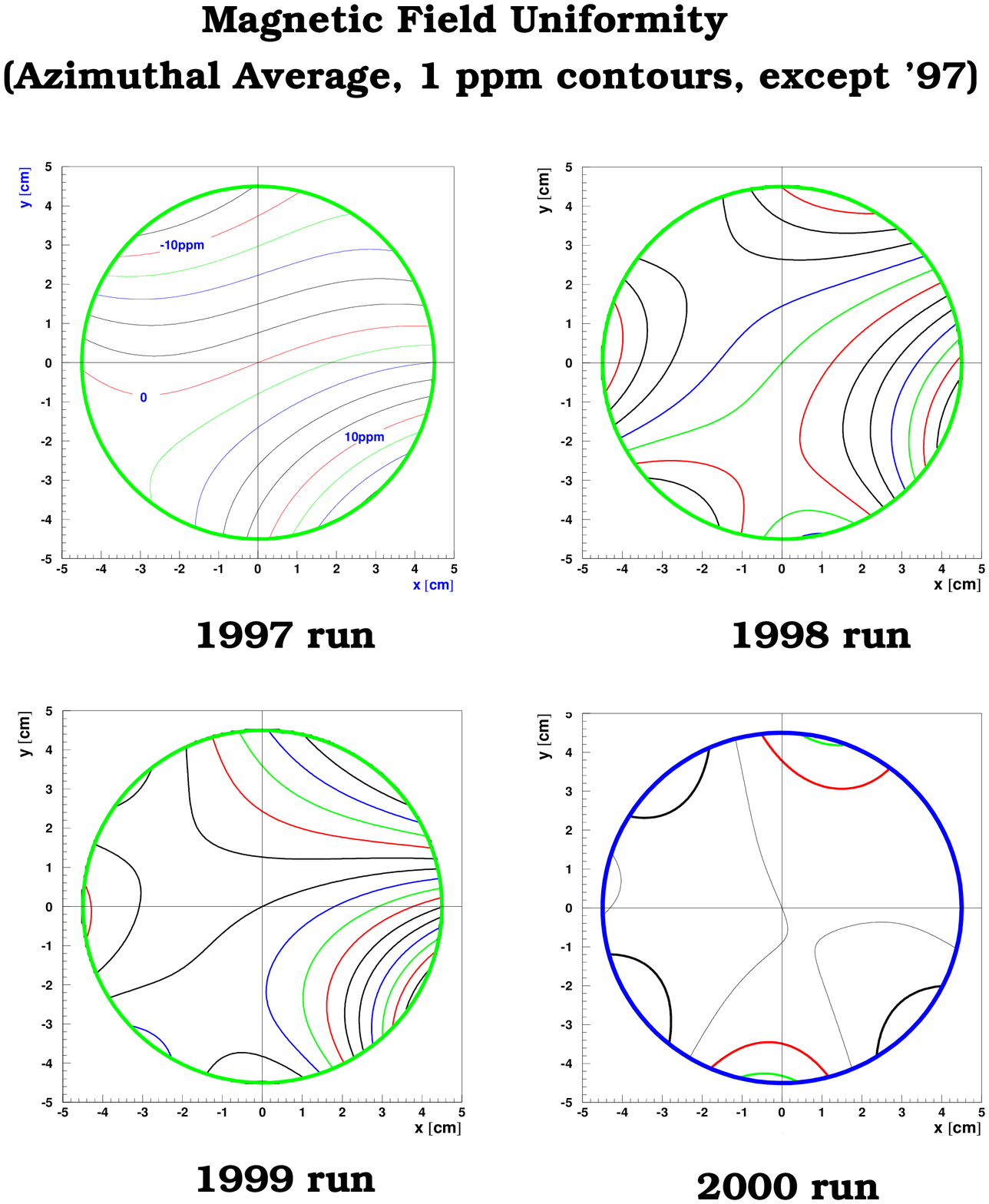}
\caption{B-field contours across the storage region in E821,
averaged in azimuth,
for succeeding run cycles. The 1997 contours
are 2 ppm, the rest 1 ppm.}
\label{fig:bf4}
\end{figure}

Many shimming options
were incorporated in order to achieve the desired field uniformity. 
The very high-quality steel
of the pole tips is decoupled from the lower
quality steel and the imperfections
(including holes for cryogenic leads, etc.) of the yoke, by means of an
air gap.
Iron wedges in the gap can be moved radially to locally adjust the
dipole field.
The thickness and position of iron pole bumps can be adjusted to minimize
quadrupole and sextupole fields. Thin sheets of iron were affixed
to the pole tips to improve local uniformity.
Current-carrying wires attached to circuit boards and mounted on the
pole faces,
with one set forming a closed loop
covering $12^0$ in azimuth and another set
going entirely around the ring, provide a final fine-tune of the dipole
field.

A continuous monitor of the B-field was provided
by 360 NMR probes placed in fixed positions around the ring,
above and below the storage region.  A subset of probes, those most highly
correlated to the average B-field, provided
feedback to the magnet power supply to compensate for the slight
field drifting which are mainly the result of ambient temperature changes.
Two separate  off-line analyses of the B-field used somewhat
different combinations of probes to determine the average field as a function
of time, with comparable results.
The B-field in the storage region was mapped in 1 cm intervals
every three to four days
with 17 NMR probes mounted transversely on a movable cable-driven trolley.
This was accomplished inside the vacuum, with essentially no geometrical
changes to the magnet or vacuum chamber configuration.
The NMR probes on the trolley were
calibrated against a standard spherical water NMR probe, which was
normalized to the precession frequency of a free proton.
The fixed probes, in the off-line analysis, tracked the
trolley probes to better than 0.15 ppm over time. The steady
improvement in the B-field provided by shimming is illustrated
in Fig.~\ref{fig:bf4}. The marked improvement from 1999
to 2000 is attributable to the replacement of the inflector,
which had a damaged superconducting fringe-field shield.

The distribution of muons inside the storage region was determined from
an analysis of the debunching of the beam as a function of time.
At the time of injection,
muons are localized in the ring with
a full width at half maximum
of about 120 degrees.
As a result, the time spectrum from a given detector at early times
will contain
oscillations with a period equal to the cyclotron period (the so-called
``fast rotation'' structure).
Muons with high momenta have smaller cyclotron frequencies than
those with low momenta, causing the bunches to spread out around the ring and
the amplitude of the oscillations to diminish
with time (``debunching'' lifetime $\approx 20 \mu s$).
The analysis of the debunching versus time
gave the radius of curvature distribution of the muons, which in
combination with simulations of the betatron motion of the muons
produces the radial and vertical distribution of muons in the storage aperture.
The  distributions thus deduced were folded geometrically
with the map of NMR frequencies to obtain $<\omega_p>$.
Corroborating information on the horizontal and vertical
distributions of muons, as well as information
on the betatron motion, at early times,
was provided by scintillating fiber hodoscopes which could be
inserted into the storage region. The hodoscopes were sufficiently thin that
useful beam profile data could be taken for many tens of
microseconds before the beam was degraded.

The final value for the average field is
${<\omega_p> \over 2\pi}=61\ 791\ 256\pm 25$ Hz (0.4 ppm).
The sources of systematic errors are given in
Table~\ref{tab:berror}. The improvements in the 2000 data set are
the installation of a new inflector
with far less fringe field, greatly reducing the first
item in the Table, and better trolley calibrations.

\begin{table}
\caption{Systematic errors in $\omega_p$ for the 1999 data set.}
\label{tab:berror}
\begin{tabular}{|c|c|} 
 
\hline 
 
\raisebox{0pt}[12pt][6pt]{Source of errors} & 
 
\raisebox{0pt}[12pt][6pt]{Error (ppm)}\\

\hline
 
\raisebox{0pt}[12pt][6pt]{Inflector fringe field} & 
 
\raisebox{0pt}[12pt][6pt]{$0.20$}\\

\raisebox{0pt}[12pt][6pt]{Fixed probe calibration} & 
 
\raisebox{0pt}[12pt][6pt]{$0.20$}\\

\raisebox{0pt}[12pt][6pt]{Fixed probe interpolation} & 
 
\raisebox{0pt}[12pt][6pt]{$0.15$}\\

\raisebox{0pt}[12pt][6pt]{Trolley $B_0$ measurements} & 
 
\raisebox{0pt}[12pt][6pt]{$0.10$}\\

\raisebox{0pt}[12pt][6pt]{$\mu$ distribution} & 
 
\raisebox{0pt}[12pt][6pt]{0.12}\\

\raisebox{0pt}[12pt][6pt]{Absolute calibration} & 
 
\raisebox{0pt}[12pt][6pt]{0.05}\\

\raisebox{0pt}[12pt][6pt]{Others$^\dag$} & 
 
\raisebox{0pt}[12pt][6pt]{0.15}\\
\hline
\raisebox{0pt}[12pt][6pt]{Total Syst error on $\omega_p$} & 
 
\raisebox{0pt}[12pt][6pt]{0.4}

\\\hline

\end{tabular}
$^\dag$ higher multipoles,
trolley temperature stability, kicker eddy
currents.

\end{table}

\subsection{Determination of $\omega_a$}

The decay positrons from $\mu^+\rightarrow e^+\nu_e \bar\nu_{\mu}$
have energies in the range $0\rightarrow 3.1 $ GeV.
In the muon rest frame,
the higher energy positrons are preferentially
emitted parallel to $\vec s_{\mu^+}$. When the muon spin is
parallel to the muon momentum, there will be more high energy muons
in the lab frame than when the directions are anti-parallel.
The number of positrons in the lab frame above a given energy threshold
$E_t$ versus time therefore oscillates at the precession frequency according
to Eq.~\ref{eq:parent},

\begin{eqnarray}
N(t) & = & N_0e^{-{t \over \tau}} \times \nonumber \\[4pt]
&&{}    (1 + A\cos{(\omega_a t + \phi_a}),
\label{eq:parent}
\end{eqnarray}
\noindent
where each of $N_0$ and $A$ 
depend strongly on $E_t$, while $\phi_a$ depends slightly on $E_t$.

The positrons, generally having lower momenta
than the muons, are swept by the B field to the
inside of the storage ring,
where they are intercepted by 24 scintillating fiber/lead
electromagnetic calorimeters\cite{sedykh} uniformly spaced around the ring.
The typical energy resolution of the calorimeters is
${\sigma (E) \over E}={8\% \over E(GeV)}$.
Since a low energy positron arrives at the detectors
more quickly after muon decay than a high energy
positron (the average distance traveled is less),
the actual measured times at the detectors relative to the
muon decay time will depend slightly on energy, therefore
$\phi_a$ depends slightly on $E_t$.
$\phi_a$ is highly correlated to $\omega_a$, implying the need to
accurately calibrate the energy scale of the calorimeters. They
were calibrated, on average,
to better than
0.2\%, using the observed energy spectra of decay positrons as a function
of time. This corresponds to an 0.02 ppm systematic error in $\omega_a$.
The average positron time measurement
was stable to 20 ps over any 200 $\mu s$ time interval,
as determined by a laser calibration system,
also giving about an 0.1 ppm systematic error.

Each calorimeter is equipped with 4 photomultiplier tubes, whose
sum is
sent to a waveform digitizer (WFD)\cite{carey} which
samples the photomultiplier pulse height every
2.5 ns. Both the time of arrival and the energy
of the positron are determined from the WFD information.

For 1999, the analysis occurred in two steps. First, in the production step,
WFD data were converted to positron energies and times.
There were two separate
productions of the data. They each developed independent algorithms
to handle the WFD data, which
eventually evolved to become similar.
The second step involved performing
$\chi^2$ minimization fits to the data 
in Fig.~\ref{fig:data} to obtain $\omega_a$.
The parent distribution in Eq.~\ref{eq:parent} provided a good $\chi^2$
fit to the
1998 data sample, with five variable parameters: $N_0$, $\tau$,
$A$, $\omega_a$ and $\phi_a$. It did not however provide a good fit to
the 1999 data, which has 15 times more positrons.
It was necessary to account for small but noticeable effects
from pulses overlapping in time at high rates (pile-up), betatron motion
of the stored muons, and muon losses.

There were four independent analyses of the positron time spectra
for the 1999 data set,
two for each production.
They used different methods to handle these additional effects.
The time spectrum, with these additional effects, can be described by
a 14 parameter function,
Eq.~\ref{eq:parent14}
(not all parameters are necessarily variable in a fit):

\begin{eqnarray}
f(t) & = & \{ N_0 e^{- { t \over \tau } }
[ 1 +A\cos{\omega_a t} 
 + 
\phi_a
] + p(t) \} \nonumber \\[4pt]
&&{}    \times  b(t)\times l(t)
\label{eq:parent14}
\end{eqnarray}

\noindent

The pile-up term, p(t), with parameters $n_p$, $A_p$, $\Delta \phi_p$,
is given by Eq.~\ref{eq:pileup}

\begin{eqnarray}
p(t) &=& N_0 e^{-2 { t \over \tau}}\nonumber\\[4pt] 
&&{}   \times ( n_p
+ A_p
\cos{(\omega_a t  + \phi_a + 
\Delta\phi_p)})\nonumber \\[4pt]  
&&{} \times ( 1 + {a_p} 
e^{ -\frac{1}{2}(\frac{t}{{\tau_p}}
)^2} )
\label{eq:pileup}
\end{eqnarray}

At times close to injection, the bunching of the muons
leads, in addition to the previously mentioned oscillations,
to an enhancement of the
pileup. The oscillations are eliminated from the time spectra by using a bin
width $=\tau_c$, adding to each arrival time a
time uniformly randomized over $\pm{\tau_c \over 2}$, and summing all
detectors around the ring.
The pile-up enhancement
is accounted for by the last term in Eq.~\ref{eq:pileup};
the constants are held
fixed to values determined in separate pile-up studies.

Two of the analyses constructed a simulated pileup time spectrum by combining
single positron pulses, from data, into pile-up pulses. The pileup spectrum
was then subtracted
from the primary spectrum, thus eliminating the p(t) term from their fits.
A third analysis varied $n_p$ and $A_p$, with
$\Delta \phi_p$ held fixed to the value determined by a
fit to an artificial pileup spectrum. Fits made with $\Delta \phi_p$ variable
produced a result consistent with other methods, but the statistical error on
$\omega_a$ was doubled as a consequence of the
strong correlations of $\Delta \phi_p$
to $\omega_a$ and $\phi_a$.

The amplitude $n_p$ of pile-up was generally less than 1\% even at the
earliest decay times,
with an asymmetry $A_p$ small compared to the (g-2) asymmetry, $A$.
The artificial pileup spectrum gave the expected ${\tau \over 2}$ lifetime,
and when subtracted from the main spectrum, did a very good job of eliminating
events above the maximum electron energy of 3.1 GeV, which apart from
energy resolution effects could only be
due to pileup. The properties of the artificial pile-up spectrum
matched very well with the results of multi-parameter pile-up fitting.

The coherent betatron oscillation (CBO) term, b(t), with
parameters $A_B$, $\omega_B$, $\phi_B$, and $\tau_B$, is described by
Eq.~\ref{eq:CBO},

\begin{equation}
b(t) = 1 + {A_B}
\cos({\omega_B} 
t 
+ {\phi_B})
\cdot
e^{- (\frac{t}{{\tau_B}}
)^2}
\label{eq:CBO}
\end{equation}

The need for b(t) is due to the effects of betatron motion
of the muons combined with the restricted aperture of the inflector.
Muons
are injected into the ring
through the inflector, whose aperture was
considerably
smaller than the storage ring aperture because of mechanical and geometrical
limitations.
This
creates a muon beam with narrow horizontal (1.8 cm)
and vertical (5.6 cm)
waists
at the inflector exit at injection time. The more important
horizontal waist case will be discussed here.
In a perfectly uniform B field
with no electric field, the muon trajectories, projected into the plane
of the magnet, are circles. After the muons are kicked, the position of the horizontal
waist would ideally be in the center of the storage region. The kick, however,
was generally less than 100\% of its optimum value.
Therefore the average radius of muons at the narrow
waist, at injection time, was larger than the central storage region radius.
At 180 degrees around the ring from the narrow waist,
the muons are spread out to fill
the ring aperture, and have an average radius more nearly equal to the
central ring radius.
The acceptance of the electron calorimeters depends to a slight extent on the
horizontal width and especially on the average radial
position of the muon beam.
Thus the detector positron acceptance will be different at the narrow waist
compared to the opposite side of the ring. 
When we add
electrostatic focusing, the position of the narrow waist (or focus)
will move
around the ring at the so-called CBO frequency, which is the
cyclotron frequency minus the horizontal betatron frequency,
$f_{CBO}=f_c(1-\sqrt{1-n})$. For the field index
$n=-{\rho \over \beta B_0}{\partial E_r \over \partial r}=0.137$,
$f_{CBO}\approx 475$ kHz.
As a result, we get a small
oscillation at $\approx 475$ kHz superimposed on the time spectrum.
The amplitude of the CBO is typically a few tenths of
a percent of the total number of counts,
and $\tau_b$ is long, about 100 $\mu$s.

One analysis allowed all four CBO parameters to vary. Two kept
$\omega_B$ fixed to the value obtained from a Fourier transform
of the residuals from a five-parameter fit. The frequencies from
the Fourier transform
and the fit were in good agreement.

The muon loss term, Eq.~\ref{eq:muloss}, has two parameters,
$a_{\mu L}$ and $\tau_{\mu L}$,

\begin{equation}
l(t) = 1+{a_{\mu L}}
\cdot
e^{- (\frac{t}{\tau_{\mu L}})}
\label{eq:muloss}
\end{equation}

Muon losses are thought to be caused by the slight drift of the orbits of
muons whose trajectories bring them close to collimators.
The drift could be caused by the small non-uniformities in the
E and B- fields,
although the exact mechanism is not known.
Indeed, when the beam is 'scraped' for about $15 \ \mu s$
right after injection,
by temporarily
displacing the stored muon beam several millimeters vertically and
horizontally in order to force the loss of muons with trajectories close
to the collimators. The rate of muon loss is markedly reduced
after the scraping is turned off, compared to the no scraping case.
After scraping, losses are generally less than $\approx$1\% at early
decay times, with the rate of losses decreasing with a short lifetime of
$\tau_{\mu L}\approx 20 \mu s$. In addition there was a roughly constant
loss rate of $\approx 0.1\%$
per muon lifetime at all decay times, as determined
by comparing the measured decay rate with
that expected from special relativity.
The two analyses which allowed both parameters to vary
obtained the same loss lifetime as was observed in a third analysis which
held $\tau_{\mu L}$
fixed to the value determined using separate muon loss detectors.

It is important to realize that only two of the 14 parameters in
Eq.~\ref{eq:parent14} have a large
correlation to $\omega_a$: $\phi_a$ and $\Delta \phi_p$. The latter
parameter received a great deal of study during the pile-up analysis.
The former introduces
no systematic error, but increases the statistical error by a factor
$\sqrt{2}$.

\begin{table}
\caption{Results of the four analyses for $\omega_a$.
$R$ is defined by $\omega_a=2\pi f_0(1-R\times 10^{-6})$.
$f_0$ is the nominal precession frequency.}
\label{tab:omresult}
\begin{tabular}{|c|c|c|} 
 
\hline 
 
\raisebox{0pt}[12pt][6pt]{\# Par.} & 
 
\raisebox{0pt}[12pt][6pt]{$\chi^2/DOF$} & 
 
\raisebox{0pt}[12pt][6pt]{R(ppm)}\\
 
\hline
 
\raisebox{0pt}[12pt][6pt]{13} & 
 
\raisebox{0pt}[12pt][6pt]{$1.012 \pm 0.023$} & 
 
\raisebox{0pt}[12pt][6pt]{$143.24\pm1.24$}\\
\cline{1-3}
\raisebox{0pt}[12pt][6pt]{10} & 
 
\raisebox{0pt}[12pt][6pt]{$1.005\pm0.023$} & 
 
\raisebox{0pt}[12pt][6pt]{$143.08\pm1.24$}\\

\cline{1-3}
\raisebox{0pt}[12pt][6pt]{9} & 
 
\raisebox{0pt}[12pt][6pt]{$1.016\pm0.015$} & 
 
\raisebox{0pt}[12pt][6pt]{$143.30\pm1.23$}\\

\cline{1-3}
\raisebox{0pt}[12pt][6pt]{3} & 
 
\raisebox{0pt}[12pt][6pt]{$0.986\pm0.025$} & 
 
\raisebox{0pt}[12pt][6pt]{$143.37\pm1.28$}\\
\cline{1-3}
\raisebox{0pt}[12pt][6pt]{Avg.} & 
 
\raisebox{0pt}[12pt][6pt]{} & 
 
\raisebox{0pt}[12pt][6pt]{$143.17\pm 1.24$}
\\\hline
\end{tabular}
\end{table}

\begin{table}
\caption{Systematic errors in $\omega_a$ for the 1999 data set.}
\label{tab:omerr}
\begin{tabular}{|c|c|} 
 
\hline 
 
\raisebox{0pt}[12pt][6pt]{Source of errors} & 
 
\raisebox{0pt}[12pt][6pt]{Error (ppm)}\\

\hline
 
\raisebox{0pt}[12pt][6pt]{Pile-up} & 
 
\raisebox{0pt}[12pt][6pt]{$0.13$}\\

\raisebox{0pt}[12pt][6pt]{AGS background} & 
 
\raisebox{0pt}[12pt][6pt]{$0.10$}\\

\raisebox{0pt}[12pt][6pt]{Lost muons} & 
 
\raisebox{0pt}[12pt][6pt]{$0.10$}\\

\raisebox{0pt}[12pt][6pt]{Timing shifts} & 
 
\raisebox{0pt}[12pt][6pt]{$0.10$}\\

\raisebox{0pt}[12pt][6pt]{E field, pitch} & 
 
\raisebox{0pt}[12pt][6pt]{0.08}\\

\raisebox{0pt}[12pt][6pt]{Binning, fit procedure} & 
 
\raisebox{0pt}[12pt][6pt]{0.07}\\

\raisebox{0pt}[12pt][6pt]{Debunching} & 
 
\raisebox{0pt}[12pt][6pt]{0.04}\\

\raisebox{0pt}[12pt][6pt]{Gain changes} & 
 
\raisebox{0pt}[12pt][6pt]{0.02}\\
\hline
\raisebox{0pt}[12pt][6pt]{Total Syst error on $\omega_a$} & 
 
\raisebox{0pt}[12pt][6pt]{0.3}\\

\\\hline
\end{tabular}
\end{table}

The fourth analysis was the novel ``ratio fit''. After pile-up subtraction,
events were
randomly assigned to four
separate time histograms. Then, the ratio in Eq.~\ref{eq:ratio} was formed:

\begin{eqnarray}
r(t) &=& {N_1^++N_2^--N_3^0-N_4^0 \over N_1^++N_2^-+N_3^0+N_4^0} \nonumber \\[4pt]
&&{} = A\cos{(\omega_at + \phi_a)}+({\tau_a \over 16 \tau_{\mu}})^2
\label{eq:ratio}
\end{eqnarray}

\noindent
where
$N_1^+=N_1(t+{\tau_a \over 2})$,
$N_2^-=N_2(t-{\tau_a \over 2})$,
$N_3^0=N_3(t)$, and
$N_4^0=N_4(t)$.
The muon lifetime cancels, and $r(t)$ is sufficiently insensitive to the CBO
and the muon losses that these effects can be neglected in the fit.
The insensitivity to the CBO is a consequence of
${2\pi \over \omega_B}$ being not so different from ${\tau_a \over 2}$.
We arrive at a three parameter fit which has different
responses to systematic errors compared to the conventional
multi-parameter fits.

The results of the four analyses are given in Table~\ref{tab:omresult}.
All of the results are well
within the bounds expected for correlated data sets.
The final value for $\omega_a$ is the average of these results,
${\omega_a \over 2\pi }=229072.8\pm0.3$ Hz(1.3 ppm), after a correction of
$+0.81\pm 0.08 $ppm for the effects of the electric field in
Eq.~\ref{eq:modomega} and for vertical
betatron oscillations (``pitch'' correction).\cite{E821b}
Note that the data are not sensitive to the sign of $\omega_a$; this however
is well-determined from many other experimental measurements, and is implicit
in the value of $\lambda$ which we use to extract the value of $a_{\mu}$. 
The sources of systematic error in $\omega_a$ are given in
Table~\ref{tab:omerr}. The AGS background is the result of 
unwanted particle injection into the ring after the initial injection.
The improvements for the 2000 data run were the addition
of a sweeper magnet in the beam-line to eliminate errors due to
the AGS background, and an
increase in the number of lost muon detectors
in order to reduce the muon loss
error.

\section{Conclusions and Outlook}

The muon anomalous magnetic moment can be both measured and calculated
(within the Standard Model) to a high precision, and given its high
sensitivity to new physics, its measurement affords an exceptional opportunity
to probe for new physics beyond the Standard Model.

The new world average value of $a_{\mu}^+$ shows a $2.2 \pm 1.4 $ ppm
difference from the Czarnecki and Marciano\cite{CM} theory compilation,
after the sign of the LOL pole term is corrected.
Many theoretical ideas have been put forward to explain any difference,
including supersymmetry, leptoquarks, muon substructure, 
etc. It could of course also be explained by a statistical fluctuation,
an error in the experiment, or an error in the Standard Model calculation.

All aspects of the theoretical calculation of $a_{\mu}$ are being heavily
scrutinized. New high quality $e^+e^-$ data from VEPP-2M and Beijing
as well as $\tau$ decay data from LEP, are being analyzed now,
and should have an impact on $a_{\mu}^{th}$ in the next few months.
Longer term, Novosibirsk and Beijing have upgrade plans,
and DA$\Phi$NE (and perhaps the B-factories and Cornell) have plans to use
the radiative return process to measure $e^+e^-$ cross sections.
Further calculations of the light-by-light term are being
considered by several groups.
One can reasonably expect a continued steady improvement in the error and
reliability of $a_{\mu}^{th}$.

In E821, analysis is under way on the $\approx$ 4 billion positrons
($\mu^+$) from the 2000 run (about four times larger than the 1999
data set)
and on the $\approx$ 3 billion electrons ($\mu^-$) from the 2001 run.
Systematic errors are expected to be reduced for both data sets.
Once these data are analyzed, it should be possible to make a more
definitive statement concerning whether the measured anomaly agrees with
theory. Comparison of the $\mu^+$ and $\mu^-$ anomalies
is a test of both the systematic errors in E821 and also CPT invariance.
With another data run, E821 expects to
achieve $0.3$ ppm statistical error on $a_{\mu}^{exp}$
and an estimated $0.3$ ppm systematic error, not far from the original goal
of 0.35 ppm overall error.

It is interesting to note that any new physics affecting $a_{\mu}$ may also
lead to a non-zero permanent electric dipole moment for the muon, through
its CP violating part. Assuming that the CP violating
phase for new physics $\phi_{CP}\approx 1$, then dimensional arguments,
along with the observed value for $\Delta a_{\mu}$,
give\cite{feng}
$d_{\mu}\approx 10^{-22}$.
Even if $a_{\mu}$ experiment and theory were to agree,
the muon EDM is interesting in
its own right: it is the only currently accessible EDM from a second
generation particle. Comparing with the electron, the current limit on the
electron dipole moment is $\approx 4 \times 10^{-27}e-cm$. If the EDM
scales by the first power of the mass, then a $10^{-24}e-cm$ muon
measurement is 
competitive with that of the electron.
There are speculations, however, that the electron EDM
could be small due to an accidental cancellation which may not apply to the
muon.
Or, if the scaling is with the square of the mass or higher, the muon then
becomes more sensitive than the electron to new physics.\cite{babu}
There are a number of models which predict $d_{\mu}$ in the range
$10^{-22}e-cm$ to $10^{-24}e-cm$.\cite{feng}

The presence of an EDM adds the term

\begin{equation}
-{e\eta \over 2mc}(\vec \beta \times \vec B + \vec E)
\label{eq:EDM}
\end{equation}

\noindent
to  Eq.~\ref{eq:modomega},
where the EDM is given
by $d_{\mu}={\eta \over 2}({e\hbar \over 2mc})$.
In the (g-2) experiment,
the effect of
the dominant $\vec \beta \times \vec B$ term is to tip the
precession vector radially by an angle
$\beta=\tan^{-1}{\eta \over 2 a_{\mu}}$.
This causes an increase in
the precession frequency to $\omega \approx \omega_a\sqrt{1+\eta^2}$.
It also causes an oscillation about zero of the
average vertical component of the positron momenta, 
which can be observed as an oscillation, with frequency $\omega$,
in the average vertical
position of positrons on the face of the calorimeters.

In the unlikely event that all of $\Delta a_{\mu}$ can be attributed
to a muon EDM, then
$d_{\mu}=(2.3 \pm 0.7) \times 10^{-19}$. 
The CERN (g-2)\cite{CERN} experiment has set the best limit on the muon EDM
so far, 
$d_{\mu} < 1\times 10^{-18}e-cm$, deduced from limits on the
vertical oscillations. While this value for $d_{\mu}$ is larger than
any theory predicts, it is nevertheless not ruled out by CERN limit.
E821 expects to reduce the limit by about a factor of five from an improved
measurement of the vertical oscillations.

A dedicated experiment to measure the muon EDM to
the $10^{-24} e-cm$ level
is currently being developed at Brookhaven National
Laboratory.\cite{dedicated}
It would use a new technique where a muon momentum and an applied
electric field would be selected so that the
second term cancels the first term in  Eq.~\ref{eq:modomega}.
One is only left with the motion in a vertical plane described by Eq.~\ref{eq:EDM}.
In this technique,
there is a very large enhancement of the EDM signal
relative to the ``noise'' over the technique used in the g-2 experiments.
It is planned to mount this experiment over the next few years.

\section{Acknowledgments}

The author would like to thank R. Carey, D. Hertzog, K. Jungmann and
Y. Semertzidis for helpful comments on this manuscript.
The author's work is supported by the U.S. National Science Foundation.
Opinions regarding the theoretical content of $a_{\mu}$ are solely those
of the author and do not necessarily reflect the opinions of the E821
experiment.
E821 is supported by the U.S. Department of Energy, the U.S. National Science
Foundation, the German Bundesminister f\"ur Bildung und Forschung, the Russian
Ministry of Science, and the U.S.-Japan Agreement in High Energy Physics.

\end{document}